\documentclass{aastex631-1}

\usepackage{placeins}
\usepackage{morefloats}
\usepackage{graphicx}

\shorttitle{Pressure balance in the LISM}
\shortauthors{Linsky and Redfield}

\begin{document}

\title{Are the Heliosphere, Very Local Interstellar Medium, and Local Cavity
in Pressure Balance with Galactic Gravity? \footnote{Based on
observations made with the NASA/ESA
Hubble Space Telescope obtained from the Data Archive at the Space
Telescope Science Institute, which is operated by the Association of
Universities for Research in Astronomy, Inc., under NASA contract NAS
AR-09525.01A. These observations are associated with programs \#12475,
12596.}}


\author{Jeffrey L. Linsky}
\affiliation{JILA, University of Colorado and NIST, Boulder, CO 
80309-0440, USA}


\author{Eberhard Moebius}
\affiliation{Space Science Center and Department of Physics, 
University of New Hampshire, 8 College Road, Durham, NH 03824, USA}
 
\correspondingauthor{Jeffrey L. Linsky}
\email{jlinsky@jila.colorado.edu}




\begin{abstract}

The {\em Voyager} spacecraft are providing the first {\em in situ} measurements of physical properties in the outer heliosphere beyond the heliopause. These data, together with data from the {\em IBEX} and {\em HST} spacecraft and physical models consistent with these data, now provide critical measurements of pressures in the heliosphere and surrounding interstellar medium. Using these data, we assemble the first comprehensive survey of total pressures inside and outside of the heliopause, in the interstellar gas surrounding the heliosphere, and in the surrounding Local Cavity to determine whether the total pressures in each region are in balance with each other and with the gravitational pressure exerted by the Galaxy. We inter-compare total pressures in each region that include thermal, non-thermal, plasma, ram, and magnetic pressure components. An important result is the role of dynamic (ram) pressure. Total pressure balance at the heliopause can only be maintained with a substantial contribution of dynamic pressure from inside. Also, total pressure balance between the outer heliosphere and pristine very local ISM (VLISM) and between the pristine VLISM and the Local Cavity requires large dynamic pressure contributions.

\end{abstract}



\section{\bf Introduction}

After \cite{Parker1958} developed a theory for the transonic outflow of the solar wind, he soon theorized how the solar wind would interact with the surrounding interstellar medium \citep{Parker1961}.  He showed that the heliosphere is structured with a termination shock, where the solar wind outflow becomes subsonic, and a heliopause, where the inflowing interstellar plasma flows around the Sun. The properties of these heliospheric regions, including their locations relative to the Sun, depend critically on the surrounding interstellar pressure. Models for astrospheres contain the same structure. These models assumed pressure balance between the heliosphere or astrosphere and interstellar plasma, but the few available measurements of these plasmas prevented a detailed study of the pressure balance. With the many available measurements from space missions and theoretical models now available, we can for the first time make a detailed assessment of the total pressures in the heliosphere and the surrounding interstellar medium to test whether pressure balance is indeed a valid assumption.

Another  critical question in studies of the local interstellar medium and its interactions with the heliosphere is whether the local interstellar environment is relatively quiescent with approximate pressure balance and small flows among its different components, or whether the environment is active with time-dependent flows that result from and/or produce large pressure imbalances among the different components. The theoretical models of the interstellar gas first proposed by \cite{Field1969} and later modified by \cite{McKee1977} and \cite{Wolfire1995,Wolfire2003}, assume time independent pressure balance among several thermal components and use this assumption to compute values of steady-state temperature and density phases that are either cold, warm (neutral or ionized) or hot. Recent detailed simulations by \cite{Gurvich2020} describe the dynamic equilibrium
and approximate balance with gravity of different thermal regimes in the disks of Milky Way-mass galaxies.

A different model emerges from the simulations by \cite{deAvillez2005} in which the energy produced by exploding supernovae produces a very active interstellar medium in which there are no stable phases but rather large variations in density, temperature, magnetic fields, and flows both spatially and temporally. The Local Bubble, also called the Local Cavity, in which the Local Interstellar Cloud (LIC) and other partially ionized warm clouds reside, was created by multiple supernova events \citep{Maiz-Apellaniz2001,Breitschwerdt2006} and could be such an active region.  However, the last nearby supernova exploded about 2.2 million years ago \citep{Breitschwerdt2016,Wallner2016}, and the interstellar gas in the initially Hot Local  Bubble (HLB)  could have cooled and settled down to a nearly quiescent state since then. 

An empirical test of these two options (or something in between) may be provided by an assessment of pressure balance or imbalance between the plasma inside and outside of the heliosheath (HS), the plasma in the outer heliosphere, the plasma outside of the heliosphere, and the Local Cavity. \cite{Jenkins2009} and others have surveyed the long standing problem of the apparent imbalance between the thermal pressures in the warm partially ionized gas surrounding the heliosphere in the LIC and other nearby clouds and the assumed million degree gas in the Local Cavity. 

\begin{figure*}[htb!]
\includegraphics[width=14.0cm,angle=-0]{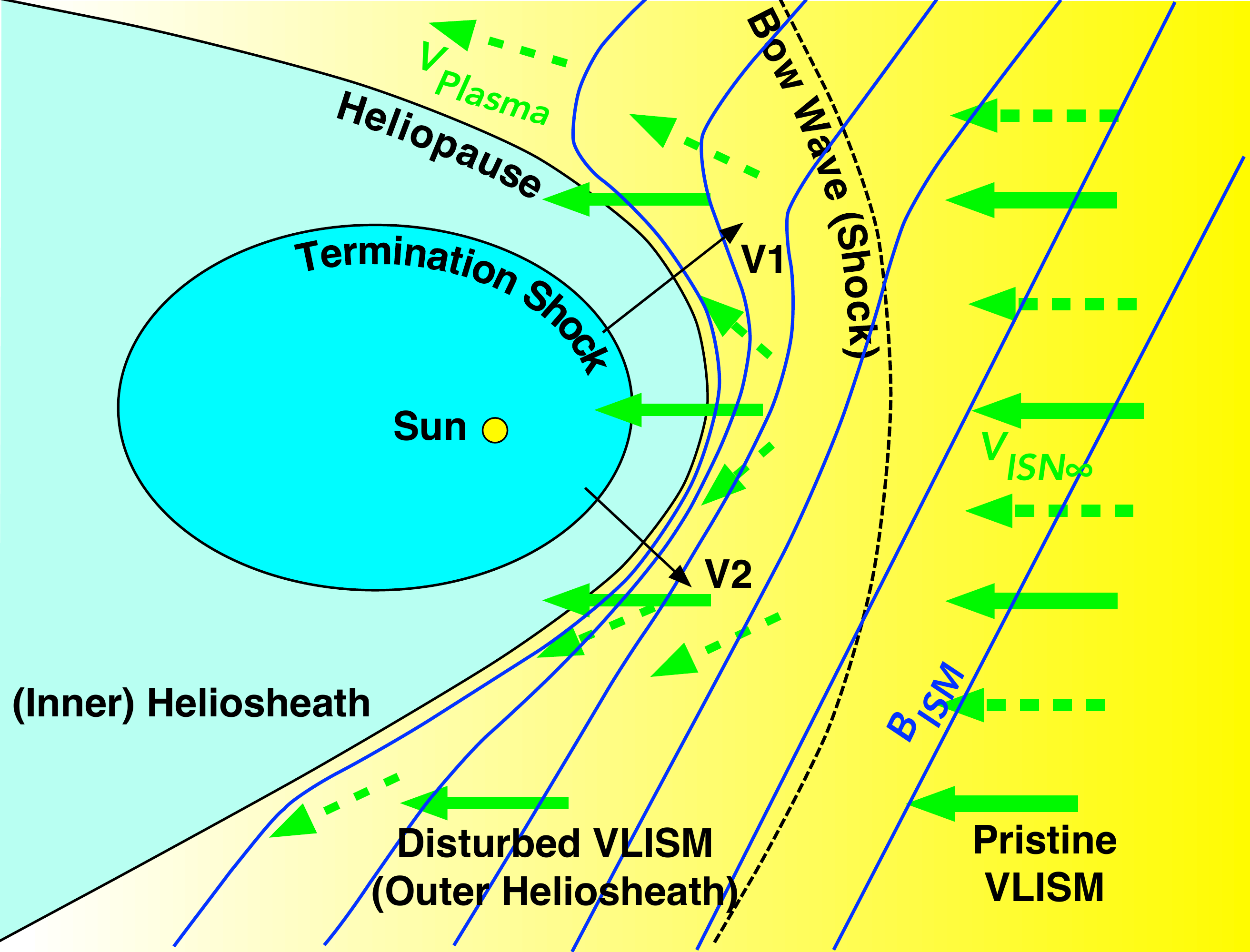}
\caption{Fig. 1: Schematic cut through heliosphere and the VLISM in the meridional plane that contains the interstellar magnetic field ({\bf B}$_{\rm ISM}$) and the Sun. The undisturbed VLISM lies outside the Bow Wave or Bow Shock on the right, where the interstellar flow is unaffected by the presence of the heliosphere. In the pristine VLISM neutral atoms (solid arrows) and plasma (dashed arrows) have the same velocity $V_{\rm ISNinfinity}$ relative to the Sun. Between the heliopause and Bow Wave is the disturbed VLISM (or Outer Heliosheath), where the interstellar plasma flow is slowed down and diverted around the HP. Here, the interstellar magnetic field ({\bf B}$_{\rm ISM}$)  is compressed and draped around the HP. The maximum compression occurs where ${\bf B}_{\rm ISM}$  is parallel to the HP. Between the HP and the TS is the Heliosheath or Inner Heliosheath, which contains the subsonic solar wind, along with pickup ions and supra-thermal particles. The HP separates the solar and interstellar domains.
Also indicated are the trajectories of {\em Voyagers} 1 and 2, projected into this meridional plane.\label{Pbalance-Fig1}}
\end{figure*}

Moving outward from the solar corona through the heliosphere toward the interstellar medium, the plasma, magnetic field, and main sources of ionization and pressure are very different on either side of shocks and magnetic separation layers. There are very different component pressures in these regions, but are the total pressures in rough pressure balance? Since papers in the literature often describe these regions and boundaries by different names, we list in Table~\ref{Regions} the terms that we will use in this paper and the boundaries of these regions and their locations. Figure~\ref{Pbalance-Fig1} provides a two-dimensional view of these regions.

\begin{table}
\begin{center}
\caption{Regions of the Heliosphere and Local Interstellar Medium}\label{Regions}
\begin{tabular}{lcc}
\hline\hline
Region or Boundary & Location (au)& Pressure terms or defining processes\\
\hline
Inner Heliosphere (IHS) & Solar Corona to TS & super-sonic solar wind (SW)\\
Termination shock (TS) & 84\tablenotemark{a}, 91\tablenotemark{b} & PUI heating and energization across the TS\\
Heliosheath (HS) & TS to HP & sub-sonic SW, supra-thermal particles\\
& &  CRs (Galactic and anomalous)\\
Heliopause (HP) & 119\tablenotemark{a}, 122\tablenotemark{b} & shocks, pile-up, magnetic separation layer\\
Disturbed Very Local ISM (disturbed VLISM) & HP to BS/BW & thermal and supra-thermal plasma, Galactic CR\\
Hydrogen wall (HW) & 200--400\tablenotemark{c} & charge-exchange processes, decelerated H~I\\
Bow shock/wave (BS/BW) & 500--700\tablenotemark{c} & Uncertain whether a shock\\
Very Local Interstellar Medium (VLISM) & Beyond the BS/BW & Beyond solar influences\\
Local Interstellar Medium (LISM) & Warm clouds within \~10 pc & Partially-ionized, closely packed within 4 pc\\
Local Bubble or Local Cavity (LB or  LC) & Beginning about 4 pc & Hot plasma or warm ionized hydrogen\\
\hline
\end{tabular}
\tablenotetext{a}{Voyager 2: \cite{Burlaga2008,Stone2019}}
\tablenotetext{b}{Voyager 1: \cite{Burlaga2005,Krimigis2013,Stone2013}}
\tablenotetext{c}{\cite{Zank2015}}
\end{center}
\end{table}

The termination shock (TS) separates the supersonic solar wind plasma from the heliosheath (HS), where the solar wind (SW) outflow is subsonic and the heated plasma contains non-thermal pick-up ions (PUIs) and suprathermal tails that are accelerated further that dominate the local pressure.
The {\em Voyager} spacecraft detected the location of the TS from changes in the SW speed at different distances from the Sun in the northern and southern heliosphere.

At a distance of about 120~au, the heliopause (HP) separates hot solar plasma in the heliosheath (HS) from much cooler interstellar plasma flowing into the heliosphere and around the HS controlled by the solar magnetic field. \cite{Zank2015} introduced the term ``very local interstellar medium (VLISM)" to indicate the region outside of the HP. 
This plasma is primarily interstellar, but the flow is slowed down, and the plasma is heated  by the injection of pick up ions.
The term VLISM, therefore, does not clearly convey the fact that interstellar plasma has been modified. Beyond the heliosphere, the interstellar gas is likely a mixture of the LIC and G clouds \citep{Swaczyna2022a}, but is unaffected by the heliospheric obstacle and is, therefore, pristine. To properly convey the physical conditions of the plasmas inside and outside of the heliosphere, we propose to use the term "disturbed VLISM" for the plasma outside of the HP where heliospheric influence is present and "pristine VLISM" further upwind of the heliosphere where heliospheric influences are no longer present. The distance at which the disturbed VLISM becomes pristine VLISM is not yet known because the question of whether the outer edge of the heliosphere is cleanly defined by a bow shock or by a more gradual bow wave is not yet settled. Within the disturbed VLISM, also called the outer heliosheath  (OHS), the charge-exchange reactions between protons and inflowing interstellar hydrogen atoms produce a region of compressed, heated, and decelerated neutral hydrogen called the hydrogen wall (HW).

It was previously thought that the heliosphere will likely exit the LIC and enter the G cloud in the direction of the stars $\alpha$ Cen A and B in less than 2000~yrs \citep{Redfield2008}, but \cite{Swaczyna2022a} has argued that the heliosphere is already in a region where the LIC and G cloud overlap.
There are no empirical data to determine where the disturbed VLISM ends and pristine VLISM begins, but theoretical models \cite[e.g.,][]{Zank2013} include a bow shock (BS) or bow wave (BW) where the inflowing supersonic or super-alfvenic plasma becomes subsonic or sub-alfvenic in the upwind direction. If this transition occurs at the BS or BW, then the 
pristine VLISM, which immediately surrounds the heliosphere, begins at about 500--700~au. Beyond the pristine VLISM, there are a number of partially-ionized warm clouds.  Between these clouds and beyond is ionized gas in the Local Cavity (LC), which is usually considered hot ($T\approx 10^6$~K) but could be warm with fully ionized hydrogen in an HII region or Str\"omgren sphere \citep{Linsky2021}. 
  
The size of the heliosphere and its three dimensional shape are controlled by the balance of total pressures between the disturbed VLISM and pristine VLISM and the total pressure balance at the HP between the HS and the disturbed VLISM. The ionization of inflowing interstellar gas also plays an important role in these pressure balances. The heliosphere is now embedded in low density interstellar gas with $n$(H~I)$\approx 0.2$~cm$^{-3}$ \citep{Slavin2008}, but it has traversed both inter-cloud and supernova remnant regions containing fully ionized hydrogen. The heliosphere may also have traversed high-density cold clouds such as the Local Leo Cold Cloud \citep{Peek2011} with densities in excess of $10^4$~cm$^{-3}$ and pressures orders of magnitude larger than at present. In the latter case, the size of the HP would have shrunk to the orbits of Jupiter or even the Earth \citep{Zank1999,Muller2006}. \cite{Linsky2022} showed that the mean density in the LIC and nearby partially ionized clouds is about $n$(H~I)=$0.10$~cm$^{-3}$, and \cite{Swaczyna2022a} proposed that the density in the immediate environment of the heliosphere is twice this density, $n$(H~I)=$0.20$~cm$^{-3}$, because the LIC and G clouds overlap in this region.
  
As the Sun journeys through the LIC, other partially ionized clouds, and fully ionized hydrogen gas in the Local Cavity, the size of the heliosphere must respond to changes in the external pressure \citep{Muller2006}. Severe reduction in the size of the heliosphere resulting from much higher density and pressure of the surrounding interstellar medium in the past and perhaps in the future would subject planets including the Earth to direct contact with the interstellar medium, high energy cosmic rays, and shocks of nearby supernova explosions with potentially catastrophic effects on life.

Properties of the heliosheath, disturbed VLISM, prestine VLISM, and Local Cavity are constrained by the weight of overlying gas, dust, and stars perpendicular to the Galactic plane \citep{Cox2005}. If this gravitational pressure is not balanced by the total dynamic pressure of these regions, there must be outflows produced by over-pressure or inflows produced by under-pressure, associated with inward and outward motion of the respective boundaries. In addition to the usual internal pressure terms (thermal, turbulent, non-thermal, cosmic ray, magnetic), an important term is the ram pressure produced by the motion of the heliosphere through the pristine VLISM and the motion of the pristine VLISM through the Local Cavity. Although estimating the many pressure terms is a formidable task, the {\em Voyager, IBEX, Cassini, and New Horizons} spacecraft have provided measurements of many important parameters, and theoretical models provide estimates of the remaining parameters. 

This study is divided as follows. In Section 2, we describe the various pressure terms that comprise the total pressure in each region and the external pressure constraint imposed by the weight of gas, dust, and stars above the Galactic plane. Sections 3 and 5 present the pressure terms inside of the HP (the heliosheath) and outside of the HP. Sections 4 and 6 describe the pressure terms in the pristine VLISM, and in the Local Cavity. Section 7 inter-compares the total pressures in these regions and considers whether or not these regions are close to total pressure balance with each other and with the external constraint. We summarize in Section 8 our results and identify uncertain parameters, in particular ram pressures. 

\section{\bf Individual pressure terms and the external constraint}

The total pressure in each region of the heliosphere and VLISM consists of several components: cosmic-ray pressure $P$(cr), magnetic pressure $P$(mag)=$B^2/8\pi$ ,where $B$ is the magnetic field strength, turbulent pressure $P$(turb)=$\rho v^2$, where $\rho$ is the density and $v$ is the turbulent velocity, thermal pressure $P$(th) =nkT, where $T$ is the temperature and $n$ is the density of all contributing particle populations, the pressure of hot (pick-up) ions $P$(hot-ions), the pressure of supra-thermal ions $P$(supra-th),  and the ram pressure $P$(ram). Not all of these components contribute significantly in each region.  Pressure has units of dynes~cm$^{-2}$ or picoPascals (pPa), where 1pPa=$10^{-11}$ dynes cm$^{-2}$. It is convenient to divide the pressure by Boltzmann's constant $k=1.38\times 10^{-16}$ erg~deg$^{-1}$, in which case $P/k=72,400$~pPa has units of Kcm$^{-3}$, and is proportional to temperature (in kelvins) times density. 

We consider the pressure components on either side of the heliopause, the heliosheath inside of the HP and the disturbed VLISM outside. The interstellar plasma flow stagnates at the HP close to the upwind direction due to the combined action of the plasmas and the magnetic fields. The maximum pressure identified with {\em IBEX} ENAs \citep{McComas2014} is offset from the stagnation region, which is close to the nose. In the maximum pressure region, there is no flow perpendicular to $B$, but still considerable flow along $B$.
The {\em Voyager} missions have provided considerable pressure data inside and outside of the HP. Estimates of the pressure terms in the pristine VLISM immediately outside of the heliosphere are obtained from measurements of neutral hydrogen and to a lesser extent neutral helium flowing into the heliosphere and their PUIs. Ram pressure plays an important role in the total pressure. Finally, we compare these pressures with two different estimates of the total pressure in the Local Cavity and with the external constraint provided by gravity.

\section{\bf Total Pressure in the Heliosheath inside of the Heliopause}

In the following we gather the component pressures in the heliosheath inside of the HP based on various {\em in-situ} observations with {\em Voyager} and ENA observations with {\em IBEX} and {\em Cassini} INCA. Table~2 contains a compilation of these pressures, starting from the cosmic rays in its first two rows.

The High Energy Telescope 2 on {\em Voyager 1} monitors Galactic Cosmic Rays (GCRs) primarily from protons with energy $E > 70$~MeV/nucleon and electrons (and positrons) with $E >15$~MeV. \cite{Cummings2016} found that protons account for about 70\% of the count rate and electrons (and positrons) account for about 25\%. They measured a count rate of 2.82/second outside of the HP and about 2.25/second inside of the HP. The corresponding energy density is $0.925\pm 0.095$ eV/cm$^3$ outside of the HP and 
$0.74\pm 0.076$ eV/cm$^3$ immediately inside the HP. Since the GCRs are mostly non-relativistic, $P$(GCRs)/$k = 5,720\pm 590$ Kcm$^{-3}$ inside the HP. 

Anomalous cosmic rays (ACRs) are thought to be interstellar neutral atoms that become ionized in the heliosheath and are picked up by the solar wind and accelerated by shocks. Both {\em Voyagers} detected ACRs beyond the termination shock \citep{Cummings2013}, but where ACRs are accelerated in the HS is uncertain. The ACR pressure is 
$P$(ACRs)/$k = 1,780\pm 180$~Kcm$^{-3}$. Both cosmic ray components provide 36.6\% of the total pressure.

The magnetic field is rather weak and varies around $0.97\pm 0.11 \mu$G throughout the heliosheath \citep{Burlaga2012}, with a pressure $P$(mag)/$k = 288\pm 120$~Kcm$^{-3}$, except for passing compression regions. Likewise, the thermal pressure of the heliosheath plasma is low. Prior to the passage of the {\em Voyagers} through the TS into the HS, it was generally assumed that the outgoing solar wind plasma would decelerate at the TS from supersonic to subsonic speeds with the kinetic energy being converted into heat. However, \cite{Richardson2008} showed that only 15\% of the available solar wind kinetic energy actually heats the plasma in the HS, which leads to $P$(th)/$k = 720 \pm 110$~Kcm$^{-3}$ based on the mean density $n_p = 0.002$~cm$^{-3}$ and temperature $T = 180,000$~K measured by \cite{Richardson2008}. The magnetic and thermal components contribute only 1.4\% and 3.5\% to the total pressure.

Most of the kinetic energy heats the PUIs, which make up about 20\% of the solar wind density at the TS. Therefore, PUIs that are accelerated near and beyond the TS dominate the total pressure in the HS \citep{Richardson2008}. The PUIs and their supra-thermal tails are not measured directly on the {\em Voyagers}, but they are accessible through ENA observations from inside the heliosphere. \cite{McComas2014} showed that the partial pressure of 0.5-4.3 keV pickup protons measured by {\em IBEX} is about 27 pdyne-AU/cm$^2$ on average, peaking at 35 pdyne-AU/cm$^2$ at the maximum pressure region and showing 30 pdyne-AU/cm$^2$ at the nose. \cite{Desai2014} argued that these PUIs observed as ENAs by {\em IBEX} are formed in the heliospheath, whereas the lower energy PUIs observed as ENAs are likely formed in the disturbed VLISM.
Adopting a HS thickness of 35~AU in the nose direction for consistency with the following values \citep{Dialynas2019}, results in the component pressure $P_{\rm PUI(IBEX)}/k = 6,260\pm 700$~Kcm$^{-3}$. For the more energetic PUIs, \cite{Dialynas2022} obtained $P$(5.2-24 keV)=$0.035\pm 0.015$ pPa based on {\em Cassini} INCA ENA observations and $P$(E$ > 28$ keV)=$0.01\pm0.002$ pPa, based on {\em Voyager} LECP observations, which are noticeably lower than the earlier values given by \cite{Krimigis2010}. Here, we adopt the most recent values, which translate into $P$(5.2-24 keV)/$k = 2,540\pm 1,100$~Kcm$^{-3}$ and $P(E > 28$ keV)/$k= 725\pm140$~Kcm$^{-3}$. These supra-thermal pressures contribute 30.5\%, 12.4\%, and 3.5\% to the total pressure, respectively.

Finally, there is a radial component of the plasma flow up to the magnetic barrier observed by {\em Voyager 2}  immediately inside of the HP, $v_R = 85\pm 10$ km~s$^{-1}$ \citep{Richardson2020, Richardson2022}. The resulting dynamic pressure in the HS, $P$(dynamic)/$k = 2,450\pm 576$ Kcm$^{-3}$, contributing another 12\% to the pressure. The total is $P$(total-HS)/$k=20,500\pm 1,600$~Kcm$^{-3}$. The mean error of the total pressure $\sigma$(total) is computed from $\sigma(total)=\sqrt{\sum{\sigma_i^2}} = 1,600$~Kcm$^{-3}$, where $\sigma_i$ are the individual errors.

\begin{table}[ht]
\begin{center}
\caption{Heliosheath pressure components in the solar and HP rest frame (Kcm$^{-3}$)}\label{HSPressure}%
\begin{tabular}{lcccc}
\hline\hline
Component & Parameter & Component Pressure (pPa) & $P/k$ (Kcm$^{-3}$) & \% of $P$(total)\\
\hline
$P$(GCRs)/$k$ & $0.74\pm0.076$ eV/cm$^3$ & $0.079\pm0.0081$ & $5,720\pm590$ & 27.9\% \\
$P$(ACRs)/$k$ & $0.23\pm0.023$ eV/cm$^3$ & $0.025\pm0.0025$  & $1,780\pm180$ & 8.7\% \\
$P$(mag)/$k$ & $1.0\pm 0.2 \mu$G & $0.004\pm 0.0017$ & $288\pm 120$ & 1.4\% \\
$P$(th)/$k$ & 180,000~K, $n_p=0.002$~cm$^{-3}$ & $0.010\pm 0.0015$ & $720\pm 110$ & 3.5\%\\
$P$(plasma(IBEX))/$k$ & $E=0.5-4.3$~keV ENA flux & $0.086\pm 0.0097$ & $6,260\pm700$ & 30.6\% \\
$P$(plasma(INCA))/$k$ & 5.2--24 keV ENA flux & $0.035\pm0.015$ & $2,540\pm1,100$ & 12.4\% \\
$P$(plasma(Voyager))/$k$ & $\geq 28$ keV ions & $0.010\pm0.002$ &  $725\pm 140$ & 3.5\% \\ 
$P$(dynamic)/$k$ & 10--150 keV ions & $0.034\pm 0.0078$ & $2,450\pm 576$ & 12.0\% \\ \hline
$P$(total-HS)/$k$\tablenotemark{a} & & & $20,500 \pm 1,600$ & 100.0\% \\
\hline
\end{tabular}
\tablenotetext{a}{Note that in the rest frame of the heliosphere, the dynamic pressure on the HP is not included in which case $P$(total-HS)/$k= 18,090\pm 1,450$.}
\end{center}
\end{table}

\cite{Schwadron2011} estimated the total pressure in the heliosheath from maps of the energetic neutral atom (ENA) flux obtained from the {\em Interstellar Boundary Explorer (IBEX)} satellite located near 1~au. The ENA flux measurements refer to the globally distributed flux after subtracting the ribbon flux and the loss of ENAs between the heliosheath and 1~au from charge exchange reactions with solar wind protons. The inferred total plasma pressure is $P$(plasma) = 1.9~pdynes~cm$^{-2}$  or $P$(plasma)/$k$= 13,800 Kcm$^{-3}$ near the nose direction, assuming a realistic heliosheath thickness of 38~au. 
The plasma pressure, $P$(plasma)/$k$ includes both thermal and nonthermal PUIs, both supra-thermal and hot ion components, and the dynamic pressure of the outflowing plasma downstream of the heliopause. This plasma pressure matches the sum of the component pressures given in lines 4-8 of Table~2 of $12,700\pm 1,430$~Kcm$^{-3}$ within error bars. Note that the estimate by \cite{Schwadron2011} does not have an error bar.

The thermal heliosheath pressure up to keV energies, as obtained from {\em IBEX} ENA flux observations at the pressure maximum \citep{McComas2014}, adjusted for a heliosheath thickness of 38~au, and the supra-thermal and hot ion pressures from INCA ENA and {\em Voyager~1} ion observations \citep{Dialynas2021}. Adding the magnetic and cosmic ray pressures (Table~2) yields a total pressure 
$P$(total-HS)/$k$=13,800+288+7,500=21,588 Kcm$^{-3}$ in agreement with the previous result.

The total pressure $P$(total)/$k = 20,500\pm 1,600$~Kcm$^{-3}$ must be matched by the total pressure outside the HP to provide pressure balance. In the following, we attempt to assess this balance near the nose of the heliosphere, where the interstellar plasma flow comes to a halt at the HP. The interstellar magnetic field leads to a pressure maximum south of the nose, however,  where the magnetic field pushes almost perpendicular to its direction into the HP \citep{McComas2014}. Because the pristine VLISM ultimately provides the outside pressure for this balance and we can obtain all needed component pressure values beyond the region of heliospheric influence from a combination of direct interstellar flow, PUIs, and ENA observations, we jump first into the pristine VLISM to see which combination of component pressures will lead to closure. Then, we will return to the disturbed VLISM just outside the HP, gather observations and augment these with modeling that captures the influence of the interaction between the heliospheric obstacle and the VLISM to test for the pressure balance. After this local exercise, we will see what the implications of the interstellar pressures are in the context of the Milky~Way galaxy. In preparation for the discussion of pressure balances with the Local Cavity in Sections 6 and 7, we added the ram pressure and total pressure of the VLISM relative to the LSR in Table~3.

\section{Pressures in the Pristine VLISM} 

As stated in the Introduction, the pristine VLISM outside the heliosphere may not feature exactly the average conditions of the LIC, which has long been thought to be the interstellar cloud that surrounds the solar system. In particular, the flow vector of interstellar material relative to Sun, as observed inside the heliosphere, appears to be noticeably different from the mean flow vector of the LIC \citep{Linsky2021}. Similarly, the density immediately outside of the heliosphere appears to be substantially higher than the mean density of the LIC \citep{Linsky2022}, perhaps due to the pristine VLISM immediately outside of the heliosphere being a mixture of the LIC and G clouds \citep{Swaczyna2022a}.

The pressure components in the pristine VLISM are compiled in Table~3 starting again with the cosmic ray pressure. Beyond the heliopause, {\em Voyager-1} measured Galactic cosmic rays (GCRs) above 3 ~MeV per nucleon with a broad maximum in the energy spectrum at 10--50 MeV per nucleon \citep{Cummings2016}. The total energy density for protons, ions, and electrons, $E/V$ = 0.83--1.02 (or $0.925\pm 0.095$) eV~cm$^{-3}$, is the range in various models that include the higher energy GCRs that {\em Voyager} could not measure. This range in $E/V$ corresponds to a non-relativistic pressure $P$(cr)/$k = 2E/(3V k) = 7,150\pm 730$~ Kcm$^{-3}$. This value of $P$(cr)/$k$ was measured where the effects of the solar magnetic field are small. Since {\em Voyager 1} and {\em Voyager 2} detected no radial gradient in the cosmic ray pressure \citep{Cummings2016, Stone2019}, we assume that $P$(cr) has the same value from the HP into the pristine VLISM. In comparison with Table~2, we see that the sum of the GCR and ACR pressures inside the HP are very close to the GCR pressure outside the HP.

While the {\em IBEX}  ribbon lies inside of the disturbed VLISM at $140^{+84}_{-38}$~au \citep{Swaczyna2016}, an analysis
of the ribbon data by \cite{Zirnstein2016} resulted in a best fit magnetic field strength $B$(VLISM)$ = 2.93\pm 0.07 \mu$G, corresponding to 
$P$(mag)/$k = 2,480\pm 120$~ Kcm$^{-3}$. Using starlight polarization data to stars within 40~pc, \cite{Frisch2015,Frisch2022} show that the interstellar magnetic field shaping the heliosphere is in agreement with the field inferred from the {\em IBEX} ribbon data and is an extended ordered field, but that magnetic field filaments exist within the disturbed VLISM. Since the LIC and G clouds overlap or merge just outside of the heliosphere  \citep{Swaczyna2022a}, the magnetic field strength in the pristine VLISM as estimated from the {\em IBEX} ribbon may be different from the magnetic fields in the LIC and G cloud.
 
To compute the particle related pressures, such as thermal, turbulent, and ram pressure, we include the plasma and neutral gas densities for both H and He assuming a He/H abundance of 0.1. In comparative simulations of heliospheric sizes in response to a wide variety of interstellar medium parameters (densities, temperatures, and speeds relative to the Sun), \cite{Muller2006} found that only when including plasma and neutral gas as well as thermal and ram pressure did the variation of the radial distance of the HP from the Sun follow a unique relation with the total pressure. This can be seen in Fig. 6 of their paper, which shows a compilation of these results, along with a power law fit to the total pressure of both components. In fact, the distance of the HP scales as 
$R$(HP) $\propto P$(total)$^{-1/2}$, i.e., inversely proportional to the square root of the pressure as expected for pressure balance. For the turbulent and ram pressure, we assume equal velocities for all species, and $P$(He)/$P$(H) $\approx 0.4$ because of the He/H mass ratio of 4. The neutral H density is taken from a recent PUI analysis using $New Horizons$ SWAP observations \citep{Swaczyna2020}, the neutral He density from He$^+$ PUIs using {\em Ulysses} SWICS \citep{Gloeckler2004}, and the H$^+$, He$^+$, and electron densities from observations of the secondary neutrals from the disturbed VLISM \citep{Bzowski2019}. These densities are listed in Table~3. It is interesting to note that the recent neutral density and plasma densities combine with the neutral He density to the canonical He/H density ratio of 0.1. 

For the temperature in the pristine VLISM, we adopt $T$(ISN) = $6,150\pm 150$~K, obtained from the interstellar neutral (ISN) He flow observations after correcting for elastic collisions in the disturbed VLISM \citep{Swaczyna2022b}. This temperature refers to the pristine VLISM in immediate contact with the heliosphere, rather than the mean temperature in the LIC ($6511\pm 2773$~K \citep{Linsky2022}) or the mixture of LIC and G cloud temperatures proposed by \cite{Swaczyna2022a}.

For the turbulent velocity, we adopt the value obtained from absorption spectroscopy of the LIC \citep{Linsky2022} with $v$(turb) =  $2.54\pm 1.18$~km~s$^{-1}$. Thus, including $P$(th) = $2,070\pm 230$~Kcm$^{-3}$ and $P$(turb) = $270\pm 180$~Kcm-$^{3}$, the total pressure in the rest frame of the VLISM amounts to $P$(total-VLISM)=$12,000\pm 1,600$ Kcm$^{-3}$, of which almost 60\% is CR pressure, 20.7\% magnetic pressure, 17.3\% thermal pressure, and turbulence only contributes 0.8\%. The total pressure falls short by $\approx 8,500$~Kcm$^{-3}$ when compared with the pressure inside the HP.  Apparently, a very important component pressure is missing to achieve the pressure balance, i.e., the ram pressure. The total ram pressure of the pristine VLISM relative to the heliosphere, $P$(ram-Sun), is listed in Table~3, along with an estimate of the effective ram pressure on the heliosphere, $P$(ram-eff), based on the magnetic and plasma pressures listed in Table~3 below.

\subsection{Estimating the ram pressure of the pristine VLISM at the HP}

The maximum possible ram pressure of the pristine VLISM relative at the Sun is based on the flow speed $25.9\pm 0-.2$~km~s$^{-1}$ of neutral He relative to the Sun measured at 1 AU by \cite{Swaczyna2021}. Adding $P$(ram-VLISM) = $28,000\pm 3,600$~Kcm$^{-3}$ to the other pressure terms would bring the total pressure in the Sun’s rest frame to $P$(total-VLISM) = $40,000\pm 5,200$~Kcm$^{-3}$, which is much too high for pressure balance. Apparently, only about 27\% of $P$(ram-Sun) shows up as an effective ram pressure $P$(ram-eff) at the HP. If instead, the ram pressure is measured from the velocity difference between the local standard of rest (LSR) and the Sun, $v$(LSR-Sun)=$18.0\pm 0.9$~km~s$^{-1}$ \citep{Frisch2015}, then $P$(total-LSR)/$k=25,520\pm 3,940$~Kcm$^{-3}$.

The effective ram pressure or the fraction of $P$(ram-VLISM) that the pristine VLISM exerts on the disturbed VLISM is the forward momentum per unit area lost by particles from the {\bf pristine} VLISM that interact with plasma in the disturbed VLISM rather than passing through unimpeded or being deflected around the heliopause with little forward momentum loss. The inflows of all interstellar charged particles (electrons, protons and ions) are diverted around the heliopause by the solar magnetic field, and thus transfer only a fraction of their forward momentum to the disturbed VLISM via the magnetic field. A portion of their forward momentum adds to the thermal plasma and magnetic pressures via density compression and heating, while the rest ends up in the diverted flow around the HP. A fraction of the inflowing hydrogen atoms charge exchange with the diverted and slowed down interstellar plasma flow adding to the momentum balance in this plasma, while the remaining fraction penetrates through the disturbed VLISM without interactions and thus do not contribute to the effective ram pressure. Since helium atoms pass through the disturbed VLISM with few interactions, their contribution to the effective ram pressure is likely minimal. The effects of He plasma and neutral atoms has recently been included in global heliospheric modeling \citep{Fraternale2021}. An assessment of the effectiveness for the pressure balance at the heliospheric boundary is underway,

The effective ram pressure can be estimated from the increased magnetic pressure, obtained from {\em Voyager} measurements, and the increased thermal plasma pressure, obtained from the model by \cite{Zank2013} listed in Table~3, $P$(mag)/$k$ + $P$(th+supra-th)/$k = 16,160\pm 3,450$~Kcm$^{-3}$. When compared with the combined magnetic and thermal pressures in the VLISM listed in Table~3, $P$(mag)/$k$ + $P$(th)/$k = 4,550\pm 256$~Kcm$^{-3}$, we obtain an estimate for the effective ram pressure as the difference between these the two values, $P$(ram-eff)/$k = 16,160-4,550= 11,610\pm 4,030$~Kcm$^{-3}$. The total pressure in the pristine VLISM, including $P$(ram-eff) is then 
$P$(total)/$k = 12,000+11,640 = 23,900\pm 4,300$~Kcm$^{-3}$.

\begin{table}
\caption{Pressure Components in the {\bf Pristine} VLISM (Kcm$^{-3}$)}\label{LICPressure}
\begin{tabular}{lccc}
\hline\hline
Component & Parameter & Component Pressure\ & \% of $P$(total-LSR)\\
\hline
$P$(cr)$\tablenotemark{a}/k$ & $0.925\pm 0.095$ eV/cm$^3$ & $7,150\pm730$ & 30.7\% \\
$P$(mag)\tablenotemark{b}/$k$ & $2.93\pm 0.07 \mu$G & $2,480\pm 120$ & 10.6\% \\
$P$(th)$\tablenotemark{c}/k$ & $T=6,150\pm 150$ K & $2,070\pm 230$ & 8.9\% \\
$P$(turb)\tablenotemark{d}/$k$ & $v$(turb)=$2.54\pm1.18$ km/s & $ 270\pm 180$ & 1.2\% \\
\hline
$P$(total-VLISM)(rest frame)/$k$ & & $12,000\pm 1,600 $ & 51.5\% \\ 
$P$(ram-VLISM)/$k$ &  $v$(LISM-Sun) = $25.9\pm 0.2$ km~s$^{-1}$ &  $28,000\pm 3,600$ &  \\
\hline
$P$(total-VLISM)/$k$ & & $40,000\pm 5,200$ &  \\
$P$(ram-eff)/$k$ & difference between inside HP and VLISM rest P &  $11,610\pm 4,030$ & \\
\hline
$P$(total-ram-eff)/$k$ & & $ 23,610\pm 4,300$ & \\
$P$(ram-LSR)/$k$ &  $v$(LISM-LSR) = $16.43\pm 3.04$ km~s$^{-1}$ &  $11,300\pm 3,500$ & 48.5\% \\
\hline
$P$(total-LSR)/$k$ & & $23,300\pm 5,500$ & 100\% \\
\hline
\end{tabular}
\tablenotetext{a}{\cite{Cummings2016}.}
\tablenotetext{b}{\cite{Zirnstein2016}.}
\tablenotetext{c}{\cite{Swaczyna2022b}.}
\tablenotetext{d}{\cite{Linsky2022}; $n_H=0.195\pm 0.033$~cm$^{-3}$ \citep{Swaczyna2020}; $n_{He}=0.015\pm 0.0015$~cm$^{-3}$ \citep{Gloeckler2004};
$n_{H^+}= 0.054\pm 0.01$~cm$^{-3}$ \citep{Bzowski2019}; $n_{He^+}=0.009\pm 0.0015$~cm$^{-3}$;  \citep{Bzowski2019}; $n_e=0.063\pm 0.01$~cm$^{-3}$  \citep{Bzowski2019}.}
\end{table}

\section{Disturbed VLISM Pressure Outside of the heliopause}

The disturbed VLISM pressure components just outside of the HP are compiled in Table~4. Some of these pressures have been directly measured by the {\em Voyager} spacecraft but others come from models. As noted in Section 4, beyond the HP {\em Voyager-1} measured Galactic cosmic rays (GCRs) above 3 MeV per nucleon with a broad maximum in the energy spectrum at 10--50 MeV per nucleon \citep{Cummings2016}. The total energy density for  protons, ions, and electrons, 
$E/V$ = 0.83-1.02 eV cm$^{-3}$, and thus the GCR pressure were actually obtained in the disturbed VLISM \citep{Cummings2016}. It is worth mentioning that about 80\% of the GCR pressure is also present inside of the HP, so that at most 20\% of the CGR pressure may contribute to shaping the HP because {\em Voyager~1} and {\em Voyager~2} detected no radial gradient in the cosmic ray pressure \citep{Cummings2016,Stone2019}. We therefore assume that $P$(cr) has the same value in the VLISM and LIC. 

\cite{Burlaga2013,Burlaga2021} presented {\em Voyager 1} magnetometer measurements extending from just inside of the HP crossing at 122~au on 2012 day 330--340 to 2020 day 172. An unexpected discovery was that the solar magnetic field measured before the HP crossing and the interstellar magnetic field measured after the HP crossing have the same direction. 
At the heliopause, the magnetic field measured by {\em Voyager~1} jumped to $4.4\pm 0.1 \mu$G \citep{Burlaga2013a}. The field then continually decreased for about 3.5 months, followed by an increase to 5.6~$\mu$G due to a shock. After the shock, the magnetic field strength decreased from $B = 4.6\pm 0.3 \mu$G in 2013, obtained from the figures in \cite{Burlaga2021}, to about $B=4.0 \mu$G at the end of 2020 (149.2~au from the Sun). This later result is an extension and re-calibration (Burlaga, private communication) of the results published by \citep{Burlaga2021}. For comparison, \cite{Zirnstein2016} estimated $B=2.93\pm 0.08 \mu$G from an analysis of {\em IBEX} ribbon for the magnetic field strength in the pristine VLISM.
Most relevant for the pressure balance just outside of the HP is the enhanced magnetic field strength measured by {\em Voyager-1} ($4.4\pm 0.1\mu$G), which is in response to the ram pressure of the VLISM.
 
To evaluate the pressure balance outside of the HP, we concentrate on the nose of the heliosphere where the interstellar plasma flow comes to a halt, and thus the effective ram pressure is responsible for the increased magnetic and thermal plasma pressure. The flow geometry and the draping of the magnetic field is shown schematically in Fig. 1 in the meridional plane that contains the interstellar magnetic field direction obtained from the IBEX Ribbon \citep{Zirnstein2016}. Adopting this geometry, {\em Voyager-1} crossed the HP where the interstellar magnetic field points largely into the HP, or north of the nose and on the left side of the heliosphere relative to the ISN flow. {\em Voyager-2} crossed the HP south of the nose and on the right side, where the interstellar field is draped almost parallel along the HP, close to the maximum pressure region. Consistent with this picture, {\em Voyager-2} measured a substantially larger magnetic field, $B = 6.8\pm 0.3 \mu$G \citep{Burlaga2019}, than {\em Voyager-1} outside the HP. This measurement is close to the maximum pressure region, \citep{McComas2014} where the magnetic pressure on the HP reaches its maximum and the nose is located between the {\em Voyager-1} and {\em Voyager-2} HP crossings. We, therefore, use the average magnetic field strength between these two values, with half the difference as a very conservative uncertainty, $B= 5.6\pm 1.2 \mu$G. Thus, the magnetic pressure outside the heliopause is 
$P$(mag)/$k=B^2/8\pi k = 9,040\pm 3,300$ Kcm$^{-3}$. It should be noted that the magnetic tension force due to the curved draping over the HP is negligible compared to the magnetic pressure. For the latter, the gradient across the HP is relevant, which stretches over $\leq 700,000$~km, whereas the local tension force depends on the curvature of $B$, with a curvature radius of $\approx 120$ au (the HP distance from the Sun) and thus is miniscule compared to the pressure.

At the HP stagnation point, the thermal pressure of the plasma $P$(th) almost completely consists of thermal pressure by H$^+$, He$^+$ and electrons, because neutrals penetrate through the HP. Given the absence of flows at the stagnation point, the sum of the thermal and magnetic pressures balance the ram pressure exerted by the pristine VLISM in the disturbed VLISM.
From the observed plasma frequency, \cite{Burlaga2021} estimates the electron density in quiescent regions past the HP as $n_e=0.11$~cm$^{-3}$. 
This density is consistent with the parameters in model 2 of \cite{Zank2013} ($n$(H$^+$) = 0.09~cm$^{-3}$ + 10\% of He$^+$ and $n$(e)=0.099~cm$^{-3}$ matching for charge neutrality). These values chosen for the VLISM parameters are closest to most recent results from observations. To be consistent with the latest ISN flow velocity relative to the Sun ($v=25.9$~km~s$^{-1}$), we adjusted the resulting thermal pressure in the \cite{Zank2013} model by a factor 1.25 to $P$(th)/$k=6,960\pm 1,040$~Kcm$^{-3}$.

The fluxes of supra-thermal particles measured by the {\em Voyagers} have dropped in excess of two orders of magnitude beyond the HP, likely making this contribution to the disturbed VLISM pressure insignificant. In the models computed by \cite{Zank2013}, the pressure at 300~au in the supra-thermal tail of the velocity distribution is small, $P$(supra-th)/$k=160\pm 20$~Kcm$^{-3}$ (Haoming Liang private comm.). \cite{Dialynas2021} measured the 40-139 keV flux of PUIs just outside the HP from {\em Voyager-1} data, but the pressure is very small, $P$(hot-ions)/$k=2.17\pm 0.31 $~Kcm$^{-3}$ (Dialynas private comm.). These partial pressures and the total pressure are listed in Table~4. With $P$(total-disturbed VLISM) = $23,310\pm 4,100$~Kcm$^{-3}$, the resulting pressure balances the one inside the HP from Table 2 within uncertainties. The magnetic field and thermal plasma contribute approximately equal amounts of the total pressure outside the HP, with 38.7\% and 29.8\%, respectively, consistent with observation of the {\em IBEX} Ribbon \citep{McComas2009}. The GCR pressure decreases by 20\% from outside to inside of the HP, where ACR pressure makes up for the decrease in the GCR pressure (see Table~2).

\begin{table}[ht]
\begin{center}
\caption{Pressures components in the rest frame of the disturbed VLISM outside of the Heliopause  (Kcm$^{-3}$)}\label{VLISMPressure [Table 4 in Section 4]}
\begin{tabular}{lccc}
\hline\hline
Component & Parameter & Component Pressure\ & \% of $P$({\bf disturbed} VLISM-HP)\\
\hline
$P$(cr)/$k$ & $0.925\pm 0.095$ eV/cm$^3$ & $7,150\pm730$ & 30.7\% \\
$P$(mag)/$k$ & $5.6\pm 1.2\mu$G & $9,040\pm 3,900$ & 38.8\% \\
$P$(th)/$k$\tablenotemark{a} & $T= 28,000$~K, $n_{\rm plasma}=0.209$~cm$^{-3}$ & $6,960\pm 1,040$ & 29.8\%\\
$P$(supra-th)/$k$ & & $160\pm 20$ & 0.7\% \\
\hline
$P$(total-DVLISM)/$k$ &  & $23,310\pm 4,100$ & 100.0\% \\
\hline
\end{tabular}
\tablenotetext{a}{Pressure with these values adjusted by x1.25.}
\end{center}
\end{table}

\section{Pressure in the Local Cavity}

Recent three dimensional models of the interstellar medium within 3~kpc of the Sun   
show a region of low absorption and thus low density extending 100--200~pc from the Sun and 
surrounded in most directions with dense clouds identified by absorption in the Na~I D and Ca~II K lines. 
The shape of this Local Cavity or Local Bubble is irregular
with a few dense clouds within 70--100~pc of the Sun and low density chimneys extending
into the halo toward the North and South Galactic poles. We use the term Local Cavity, which does not indicate a temperature for the plasma, rather than Local Bubble, which usually implies that the plasma is hot, or Local Hot Bubble which clearly indicates a hot plasma. The models presented by
\citet{Capitanio2017}, \citet{Lallement2019}, and \citet{Leike2020} are based on reddening and color excess obtained from a variety of sources
including diffuse interstellar absorption bands with distances to stars from {\em GAIA}. These models describe the morphology
of the low density region surrounding the Sun. Located within the Local Cavity is the heliosphere and  
warm (5,000--10,000~K) partially ionized clouds extending 5--10~pc  outward from the Sun
\citep{Redfield2008,Frisch2011}. 

\citet{Fuchs2006} and \citet{Benitez2002} presented a convincing case that the Local Cavity was produced by 
supernova explosion blast waves that heated and evacuated the surrounding interstellar gas and produced an exterior 
dense shell of cooler gas. \citet{Breitschwerdt2016} found that a total of 14--20 supernovae over the past 13 Myr in the
Scorpius-Centaurus Association created this multiple supernova remnant with the two most recent
supernovae occurring about 2.3 Myr ago at a distance of 90--100~pc. The recent age of these two supernovae has been inferred from the presence of 
the radioactive $^{60}$Fe isotope produced by electron-capture supernovae and found embedded
in deep ocean crust samples \citep[e.g.,][]{Wallner2016}. The effect of supernova blast waves is to produce a remnant consisting of highly ionized 
million degree gas that cools by radiation, expansion, and shock heating of denser material at the edge of expansion. 
The Local Cavity was likely created by the cumulative heating, expansion, and subsequent cooling of many supernova events. The most recent of these supernovae
would have evolved inside of the Local Cavity producing a hot bubble that filled a portion or all of the the present volume of the Local Cavity. 
\citet{Shelton1999} computed the long term evolution of a supernova explosion expanding into a previously evacuated  low density (0.01~cm$^{-3}$) modest temperature
($10^4$~K) cavity. These hydrodynamic simulations that include non-equilibrium ionization could provide an approximate model for the present day Local Cavity after the most recent supernova explosions. 

After more than 40 years of intensive studies, the question of what fills the Local Cavity remains unanswered. The presence of some million degree gas is universally accepted, but
much or most of the Local Cavity could be filled with something else. Until now what fills the Local Cavity has been studied by modeling the observed diffuse X-ray emission, where it is formed, and whether it is primarily thermal emission from diffuse hot gas or is largely local emission produced when the solar wind ions charge exchange with neutral hydrogen in the heliosphere \citep{Cravens2001}. Unfortunately, the identification of the matter filling the Local Cavity is frustrated by two uncertain but critical parameters, 
the collisional excitation rates for the charge-exchange processes and the electron density in the Local Cavity. We consider here two models: the Local Hot Bubble model in which million degree gas fills the entire bubble, and a moderate temperature Str\"omgren Sphere Model \citep{Linsky2021} in which the plasma has cooled and hydrogen is fully ionized by the EUV radiation of hot stars. The very different total pressures in these two models provides an interesting test of whether the Local Cavity has remained hot or has cooled since the last supernova event.

For both models we assume the cosmic ray pressure $P$(\rm cr)/$k=7,150\pm730$ measured outside of the HP by \cite{Cummings2016}.
There are only indirect estimates of the magnetic field strength in the
Local Cavity. The \citet{deAvillez2005} simulations show a mean magnetic
pressure $P_{\rm mag}=5,580$~Kcm$^{-3}$ corresponding to an average total magnetic field
$B= 4.4 \mu$G, but the local magnetic field strengths in this
simulation have a wide range. The analysis of dispersion measures and
rotation measures of four pulsars within 300 pc of the Sun in the third Galactic quadrant,
yields $B \approx 3.3 \mu$G with a large reduced $\chi^2=40$
\citep{Salvati2010}. For longer path lengths through the Galactic plane, \citet{Sobey2019} 
derived a mean longitudinal magnetic field of $4.0\pm 0.3$$\mu$G from pulsar data.
Considering all of these values, we estimate the Local Cavity mean magnetic field strength
to be  $B=3.5\pm 0.5\mu$G, corresponding to $P_{\rm mag}=3,530\pm 1,000$~Kcm$^{-3}$.

Turbulence in supernova remnants is produced at large scales by supernova shocks and then converted to smaller scales
by interactions with density and magnetic field inhomogeneities. On intermediate scales turbulence can be generated by many processes including 
thermal instabilities, thermal shell instabilities, density inhomogeneities, and magnetic instabilities as described by \citet{Raymond2020}.
Given this complexity and the range of scales involved, there is no simple way of quantifying the turbulent pressure. \cite{Linsky2021}  proposed that 
the random motions of nearby warm interstellar clouds relative to their common velocity vector provides a rough estimate of the 
macroscopic turbulent pressure in the Local Cavity.  
The mean value of these random motions is $v=16.9$~km~s$^{-1}$ \citep{Linsky2008b,Frisch2011}.
This velocity is consistent with the 15--21 km~s$^{-1}$ rms velocities for moderate temperature gas in the \citet{deAvillez2005} simulation (see below).
Assuming that these random motions are typical of random motions within the
Local Cavity, we compute $P_{\rm turb}=\rho v^2 = 1.1n_{\rm H} m_{\rm H} v^2 =
8,610\pm 1,200$~Kcm$^{-3}$. 

The gas temperature in Str\"omgren spheres is typically
10,000--20,000~K, and the pulsar dispersion measured electron density in the Local Cavity
is $n_e=0.012$~cm$^{-3}$. If we assume a temperature $T=15,000\pm 5,000$~K, then
$P_{\rm th}/k= 2.2n_e T =330 \pm 110$~Kcm$^{-3}$.

We estimate the thermal pressure in the Local Hot Bubble Model from the temperature $T=(1.18\pm 0.01)\times10^6$~K \citep{Snowden2014} and electron density $n_e=0.0121\pm0.0029$, based on pulsar dispersion measurements summarized in \cite{Linsky2021}. With $n_{\rm total}=1.92n_e$, the resulting 
thermal pressure is $P_{\rm th}/k=27,100\pm 1,780$~Kcm$^{-3}$. 
\cite{Snowden2014}, on the other hand, estimated the thermal pressure in the Local Bubble from the X-ray emission after correction for local charge-exchange component. Their result is $P$(th)/$k=10,700$~Kcm$^{-3}$, but this result is based on an electron density $n_e=0.00468\pm0.00047$, which is a factor of 2.6 smaller than the Pulsar dispersion value. We therefore consider this estimate of the thermal pressure to be too low.
The sum of the four pressure terms is then $19,620\pm 1,730$~Kcm$^{-3}$ for the Str\"omgren Sphere Model  but $46,390\pm2,480$~Kcm$^{-3}$ for the Local Hot Bubble model. The pressure components and total pressures for both models are shown in Table~\ref{tab:LCpressures}. 

\begin{table}
   \caption{Components of the Total Pressure (Kcm$^{-3}$) in the Local Cavity rest frame}
    \label{tab:LCpressures}
\begin{center}
\begin{tabular}{lccc}
  \hline\hline
Component & Parameter & Str\"omgren Sphere Pressure & Hot Bubble Pressure\\  
\hline
$P_{\rm cr}/k$ & $0.925\pm 0.095$ eV/cm$^3$ & $7,150\pm730$ & $7,150\pm730$\\
$P_{\rm mag}/k$ & $B=3.5\pm 0.5~\mu$G & $3,530\pm1,000$ & $3,530\pm 1,000$\\ 
$P_{\rm turb}/k$ & $v=16.9$ km~s$^{-1}$ & $8,610\pm 1,200$ & $8,610\pm 1,200$\\
Sum & & $19,290\pm 1,730$ & $19,290\pm 1,730$\\
$P_{\rm th}/k$ & $T$$=15,000\pm 5,000$~K & $330\pm 110$ & \\
                     & $T=(1.18\pm 0.01)\times10^6$~K & & $27,100\pm 1,780$\\
\hline
$P_{\rm total}/k$ & & $19,620\pm 1,730$ & $46,390\pm 2,480$\\
\hline\hline
\end{tabular}
\end{center}
\end{table}

 \section{Are the total Pressures in the Heliosphere, pristine VLISM, and Local Cavity in Equilibrium with each other and with the Galactic Gravitational Pressure?}

In Table~\ref{Pressurecompare} we list pressures in the rest frame ($2^{\rm nd}$ column) of the respective region and where appropriate including ram pressure in the applicable frame of the surroundings ($2^{\rm nd}$ and $3^{\rm rd}$ column). We compare the total pressures in the heliosheath just inside, just outside of the HP, the pristine VLISM, and the Local Cavity to determine whether there may be significant imbalances that would cause relative flows. Within their uncertainties, the total pressure outside the heliopause ($23,310\pm4,100$ Kcm$^{-3}$) is consistent with the pressure inside ($20.500\pm1,600$~Kcm$^{-3}$) including the effective ram pressure relative to the Sun. However, other presently unknown pressure sources could be present. The total pressure of the pristine VLISM including ram pressure is consistent with the total pressure in the disturbed VLISM within the uncertainties. These conditions indicate pressure balance among these regions and no anticipated flows or motion of the boundaries. To reiterate, the motion of the pristine VLISM relative to the Sun and thus its ram pressure is essential to providing pressure balance at the nose of the heliosphere. In addition to the pristine VLISM pressures, we add the LIC here as the closest pristine interstellar cloud, because according to the recent findings by \cite{Swaczyna2022a}, the locally accessible pristine VLISM is very likely an interaction region between the LIC and the G-Cloud and thus with higher densities and pressures than the individual clouds. We use 50\% of all densities in the pristine VLISM for the LIC and temperatures and velocities from absorption line observations \citep{Redfield2008}. We keep the magnetic field strength although it may be somewhat compressed in the pristine VLISM.

The comparison of the total pressure in the pristine VLISM with that of the Local Cavity depends on which model for the Local Cavity one assumes. For the Str\"omgren Sphere model, the total pressures, including ram pressure are comparable,
but for the Hot Local Bubble model there is a severe imbalance as the LHB model pressure is more than twice as large as in the pristine VLISM, implying the need for significant flows to balance the pressure difference. Even if one were to accept the thermal pressure of 10,700~Kcm$^{-3}$ estimated by \cite{Snowden2014}, the total pressure for the LHB model is reduced only to 29,890~Kcm$^{-3}$, which is still much larger than the pristine VLISM. Considering that substantial ram pressure is needed for the pressure balance, which would leave the wake of the clouds exposed to the surrounding pressure of the Local Cavity, it is apparent that, with a bulk speed of 10-20 km~s$^{-1}$ the clouds would not be able to outrun the thermal speed of $\approx 310$~km~s$^{-1}$ in the Hot Bubble Model, while the bulk speed is comparable in the Stro\"mgren Sphere model and would still maintain a wake keeping the clouds also from major inflows on their rear end. The high pressure and large thermal speed are 
arguments that the LHB model is unrealistic and that the Str\"omgren Sphere model should be accepted.

Finally, let us compare the pressures in these regions with the weight of overlying material perpendicular to the Galactic plane, as estimated by \cite{Cox2005}. Since the Sun is very close to the Galactic plane, the gravitational pressure is $P$(grav)=$3.0\times 10^{-12}$ dynes~cm$^{-2}$ or $P$(grav)/$k=22,000$~Kcm$^{-3}$. We call attention to the remarkable near agreement between the gravitational pressure and the total pressures the heliosphere (inside and outside of the HP), the pristine VLISM, and the Local Cavity, except for the LHB  model. For the disturbed VLISM and pristine VLISM, the pressure is also approximately in agreement when including the total ram pressure for motion relative to the Local Standard of Rest (LSR) based on \cite{Frisch2011}, which may be considered as the appropriate rest frame for the wider neighborhood of the Sun on its orbit around the galactic center. However, when only using the pressures in the rest frame of the specific cloud they fall short by approximately a factor of two. 

 Is our result that the total pressures in the heliosheath, including appropriate ram pressures in the heliosheath, disturbed VLISM,
and surrounding interstellar gas are in approximate balance with the weight of material above the disk in agreement with recent simulations for galactic disks? \cite{Gurvich2020} used the FIRE-2 galaxy simulation code to study the structures and properties of the multiphase ISM in disks of galaxies with mass typical of the Milky Way. Their simulations included thermal, turbulent, and dynamic pressures as a function of radial distance from the center and height above the disk. The resulting calculations for the disk midplane show that total pressures in the dynamic ISM are typically between 80\% and 100\% of the weight of overlying material, consistent with our results.

The pressure in the Local Cavity could be far from balance with the gravitational pressure for a number of reasons. One is that the internal velocities created by the last supernova explosion about 2 Myr ago may still be present as shocks producing higher pressures at them or shock lower pressures in their wake. A second reason is that hot gas, if present, would have high thermal pressure leading to expansion of the gas toward the Galactic poles as is observed. Finally, the Local Cavity may still be expanding into the surrounding medium in which case there would be a rarefaction inside the cavity with lower pressure, similar to the rarefaction region in the solar wind behind coronal mass ejections and the compressions behind stream interaction regions \citep{Pizzo1978}. Conversely, a ram pressure term that should be included in the total pressure would raise the pressure just outside the Local Cavity.

\begin{table}[h]
\begin{center}
\caption{Comparison of Total Pressures Components in Different Regions}\label{Pressurecompare}%
\begin{tabular}{llll}
\hline\hline
Component & \multicolumn{3}{c}{Total Pressure (Kcm$^{-3}$)}\\
                    & in its rest frame & including P(ram) & including P(ram)\\
                    &                           &relative to Sun or HP     &relative to LSR \\
\hline
Heliosheath inside of the HP &  $18,050\pm 1,450$ &  $20,500\pm 1,600$ & \\
Disturbed VLISM outside of the HP & $23,310\pm 4,100$ & & \\
Pristine VLISM & $12,000\pm 1,600$ &  $23,610\pm 4,300$ &  $23,300\pm 5,500$\\
Str\"omgren Sphere model of the LC & $19,620\pm1,730$ & & \\
Local Hot Bubble model of the LC & $46,390\pm2,480$ & & \\
Galactic gravitational pressure & 22,000 & &\\ 
\hline
\end{tabular}
\end{center}
\end{table}

\section{Conclusions}

     The question of pressure balance or imbalance between the heliosphere and the surrounding interstellar medium provides important insight into whether the local region of space is relatively inactive with weak flows among the regions or is highly dynamic with strong flows indicative of a young supernova remnant. To test for total pressure balance, we assembled the first comprehensive study of pressures in the heliosphere (inside and outside of the heliopause), the pristine VLISM, and the surrounding Local Cavity produced by a series of supernova explosions. To test the validity of total pressure balance, we cited or computed the thermal, non-thermal, plasma, ram, and magnetic pressure components in these regions based on {\em Voyager, IBEX, Ulysses, New Horizons,} and {\em HST} measurements and models consistent with these measurements with the following results:
     
\begin{itemize}

\item In the heliosheath, the region inside of the heliopause and outside of the temination shock, the pressure of 0.7-24~keV ions and electrons dominates the total pressure, although cosmic rays (Galactic and anomalous) also contribute. To balance the total pressure outside of the heliopause, it is essential to include the dynamic pressure that the heliosheath flow
exerts on the gas just outside of the heliopause.

\item Outside of the heliopause, in the region called the disturbed very local interstellar medium (disturbed VLISM)  or the outer heliosheath (OHS), the cosmic ray, magnetic, and plasma pressures contribute equally to the total pressure. The sum of the magnetic and plasma pressures in the stagnation region just outside of the heliopause balance the ram pressure of plasma inflowing from the LIC resulting from the heliosphere's motion through the interstellar medium. The total pressures in the heliosheath ($20,500\pm 1,600$~Kcm$^{-3}$) and outside of the heliopause ($23,310\pm 4,100$~Kcm$^{-3}$) are in agreement within their respective errors.

\item In the pristine VLISM the pressure components of Galactic cosmic rays, magnetic fields and thermal pressure only sum to $12,000\pm 1,600$~Kcm$^{-3}$, which is far below the total pressure in the heliosphere. What is missing is the ram pressure produced by the motion of the pristine VLISM relative to its environment. The maximum possible ram pressure would be due to the $25.9\pm 0.2$~km~s$^{-1}$ speed of the Sun through the pristine VLISM, as measured by the flow of neutral helium into the heliosphere. However, ions and electrons flow around the heliosphere and transfer only a fraction of their forward momentum to the disturbed VLISM. Neutral hydrogen mostly flows through the disturbed VLISM transferring only a portion of its forward momentum to the plasma in the disturbed VLISM, and neutral helium flows through with almost no momentum transfer. Although detailed simulations of these effects are underway, we estimate the effective ram pressure from the difference between the sum of magnetic and thermal pressures in the disturbed VLISM and the corresponding sum in the pristine VLISM. The difference is produced by the compression and heating of the disturbed VLISM plasma and magnetic field by the ram pressure. The effective ram pressure is then $11,610\pm 4,030$~Kcm$^{-3}$, and the resulting total pressure in the pristine VLISM is $23,610\pm 4,300$~Kcm$^{-3}$. An estimate of the ram pressure relative to the local standard of rest results in a very similar total pressure ($23,300\pm 5,500$~Kcm$^{-3}$). Compared with the other component pressures in the disturbed VLISM that contribute to the pressure balance at the heliopause, the effective ram pressure dominates of the magnetic field and thermal pressures, exceeding their combination by a factor of 2.0.

\item The pressure of Galactic cosmic rays, magnetic field, and turbulent motions sum to a total pressure of $19,290\pm 1,730$~Kcm$^{-3}$ in the Local Cavity. The additional contribution of thermal pressure depends on the assumed model. Inclusion of the thermal pressure of million degree gas for the Local Hot Bubble model raises the total pressure to $46,390\pm 2,480$~Kcm$^{-3}$, although the total pressure would be only $27,400\pm6,530$ if the electron density is as low as 0.0047~cm$^{-3}$. However, the warm gas in the Str\"omgren sphere model results in a total pressure of $19,620\pm 1,730$~Kcm$^{-3}$. We believe that this model is more realistic on the basis that (1)  the total pressure is consistent with that in the heliosphere, and the pristine VLISM, when including the ram pressures in the applicable rest frames. (2) The pressures are comparable with gravitational pressure (22,000~Kcm$^{-3}$). Future work should include the presence of some hot gas and possible non-thermal pressures.

\item This first survey of the total pressures in the heliosphere, pristine VLISM, and Local Cavity, leads to a scenario in which all of these regions are close to pressure balance with each other and with the 22,000~Kcm$^{-3}$ gravitational pressure due to the weight of gas above and below the Galactic plane. The inclusion of ram pressure is essential for computing realistic total pressures. Uncertainties remain, in particular possible missing pressure terms, more realistic ram pressures, and the uncertain electron density in the Local Cavity.

\end{itemize}

Overall, it is interesting to note the approximate balance of the 
total pressures of the warm interstellar clouds in the solar neighborhood (including dynamic pressures
of their relative motions),  the surrounding Local Cavity, and the gravitational pressure of the gas on the Galactic plane.
This balance even extends to the total pressure of the pristine VLISM on the heliosphere. At this point, we can only
speculate concerning the reason(s)  for this overall approximate balance, which may depend upon the distribution of the gravitational forces among the
motion of the stars and the interstellar gas, along with their internal pressures. A discussion of this topic is beyond the scope of this paper
and may be taken up in a future investigation.\\

Facilities: {\em HST}(STIS), {\em HST}(HRS),  {\em EUVE},  {\em Voyager I, Voyager II},  {\em IBEX},  {\em Ulysses},  {\em New Horizons}\\

Acknowledgements: J. L. thanks NASA for support through grant No. 80NSSC20K0785 to Wesleyan University and the University of Colorado. E. M. gratefully acknowledges support from the NASA IBEX and IMAP missions and NASA Grant 80NSSC18K1212. E. M. thanks D. McComas and N. Schwadron for insightful discussions on the processes at the heliopause. J. L. thanks P. C. Frisch for calling attention to her value of the Sun's velocity relative to the Local Standard of Rest. We thank the International Space Science Institute for the hosting the Workshop on the Heliosphere and the Local Interstellar Medium where the idea for this paper was conceived.


\begin{thebibliography}{}
  
\bibitem[Benitez et al.(2002)]{Benitez2002} Benitez, N., Maiz-Apellaniz, J., Canelles, M.\ 2002, Phys. Rev. Lett., 88, 081101

\bibitem[Breitschwardt \& de Avillez(2006)]{Breitschwerdt2006} Breitschwerdt, D., de Avillez, M.A.\ 2006, \aap, 452, L1
 
\bibitem[Breitschwerdt  et al.(2016)]{Breitschwerdt2016} Breitschwerdt, D., Feige, J., Schulreich, M.~M., de Avillez, M.~A., Dettbarn, C., Fuch, B.\ 2016, Nature, 532, 73
 
\bibitem[Burlaga et al.(2021)]{Burlaga2021}  Burlaga,  L.~F., Kurth,  W.~S., Gurnett,  D.~A., Berdichevsky,  D.~B., Jian,  L.~K., Ness, N.~F., Park, J., Szabo, A.\ 2021,  \apj, 911, 61

\bibitem[Burlaga \& Ness.(2012)]{Burlaga2012} Burlaga, L.F., Ness, N.~F.\  2012, \apj, 744, 51

\bibitem[Burlaga et al.(2005)]{Burlaga2005} Burlaga, L.F., Ness, N.~F., Acu\~na, M.~H., et al.\ 2005, Science, 309, 2027
 
\bibitem[Burlaga et al.(2008)]{Burlaga2008} Burlaga, L.F., Ness, N.~F., Acu\~na, M.~H., et al.\ 2008, Nature, 454, 75

\bibitem[Burlaga et al.(2019)]{Burlaga2019} Burlaga, L.~F., Ness, N.~F., Berdichevsky, D.~B.\ 2019, Nat. Astronomy, 3, 1007

\bibitem[Burlaga et al.(2013)]{Burlaga2013} Burlaga, L.~F., Ness, N.~F.,  Gurnett, D.~A., Kurth, W.~S.\ 2013,  \apj, 778, L3

\bibitem[Burlaga, Ness, \& Stone(2013)]{Burlaga2013a} Burlaga, L.~F., Ness, N.~F., Stone, E.~C.\ 2013, Science, 341, 147

\bibitem[Bzowski et al.(2019)]{Bzowski2019} Bzowski, M., Czechowski, A., Frisch, P.~C.\ 2019, \apj, 882, 60

\bibitem[Capitanio et al.(2017)]{Capitanio2017} Capitanio, L., Lallement, R.,
Vergely, J.~L., Elyajouri, M., Monreal-Ibero, A.\ 2017, \aap. 606, A65

\bibitem[Cox(2005)]{Cox2005} Cox, D.~P.\ 2005, \araa, 43, 337
 
\bibitem[Cox \& Snowden(1986)]{Cox1986} Cox, D.~P., Snowden, S.~L.\ 1986, Adv. Space. Res., 6, 97

\bibitem[Cravens, Robertson \& Snowden(2001)]{Cravens2001} Cravens, T.~E., 
Robertson, I.~P., Snowden, S.~L.\ 2001, \jgr, 106, 24883

\bibitem[Cummings et al.(2013)]{Cummings2013} Cummings, A.~C., Stone, E.~C.\ 2013, AIP Conf. Proc. 1516, 97

\bibitem[Cummings et al.(2016)]{Cummings2016}  Cummings,  A.~C., Stone,  E.~C., Heikkila,  B.~C., Lal, N., Webber,  W.~R., Johannesson,  G.,  Moskalenko, I.~V., Orlando,  E., Porter, T.~A.\ 2016,  \apj,  831 18
 
\bibitem[de Avillez \& Breitschwerdt(2005)]{deAvillez2005} de Avillez, M.~A. \& Breitschwerdt, D.\ 2005, \aap, 436, 585

\bibitem[Desai et al.(2014)]{Desai2014} Desai, M.~I., Allegrini, F.~A., Bzowski, M., et al.\ 2014, \apj, 780, 98
 
 \bibitem[Dialynas et al.(2022)]{Dialynas2022} Dialynas, K., Krimigis, S.~M., Decker, R.~B., Hill, M.,  Mitchell, D.~G.  Hsieh, K.~C., Hilchenbach, M., Czechowski, A.\ 2022,  \ssr, 218,  21D
 
\bibitem[Dialynas et al.(2019)]{Dialynas2019} Dialynas, K., Krimigis, S.~M., Decker, R.~B., Mitchell, D.~G.\ 2019, GeoRL, 46, 7911
 
\bibitem[Dialynas et al.(2021)]{Dialynas2021} Dialynas, K., Stamatios, K., Decker, R.~B., Hill, M.~E.\ 2021,  \apj,  917, 42
  
\bibitem[Field, Goldsmith, \& Habing(1969)]{Field1969} Field, G.~B., 
Goldsmith, D.~W., \& Habing, H.~J.\ 1969, \apjl, 155, L149

\bibitem[Fraternale et al.(2021)]{Fraternale2021}  Fraternale, F.,  Pogorelov, N.~V., Heerikhuisen, J.\ 2021, \apj, 921, L24

\bibitem[Frisch et al.(2015)]{Frisch2015} Frisch, P.~C., Berdyugin, A., Piirola, V., et al.\ 2015, \apj, 814, 112

\bibitem[Frisch, Redfield, \& Slavin(2011)]{Frisch2011} Frisch, P.,C., Redfield, S. \& Slavin, J.~D.\ 2011, \araa, 49, 237

\bibitem[Frisch et al.(2022)]{Frisch2022} Frisch, P.~C., Piirola, V., Berdyugin, A.~B., et al.\ 2022, \apjs, 259, 48  

\bibitem[Fuchs et al. (2006)]{Fuchs2006}  Fuchs, B., Breitschwerdt, D., De Avillez, M.~A., Dettbarn, C., Flynn, C\ 2006, \mnras, 373, 993

\bibitem[Gloeckler et al.(2004)]{Gloeckler2004} Gloeckler, G., M\"obius, E., Geiss, J. Bzowski, M., et al. \ 2004,  \aap, 426, 845

\bibitem[Gurvich et al.(2020)]{Gurvich2020} Gurvich, A.~B., Faucher-Gigu\`ere, C.-A., Richings, A.~J., et al.\ 2020, \mnras, 498, 3664

\bibitem[Jenkins(2009)]{Jenkins2009} Jenkins, E.~B.\ 2009, \ssr, 143, 2005

\bibitem[Krimigis et al.(2013)]{Krimigis2013} Krimigis, S.M., Decker, R.~B., Roelof, E.~C., Hill, M.~E., Armstrong, T.~P., Gloeckler, G., Hamilton, D.~C., Lanzerotti, L.J.\ 2013,  Science, 341, 144
 
\bibitem[Krimigis et al.(2010)]{Krimigis2010} Krimigis, S.~M., Mitchell, D.~G., Roelof, E.~C., Decker, R.~B.\ 2010, AIP Conf. Proc.,1302, 79

\bibitem[Lallement et al.(2019)]{Lallement2019} Lallement, R., Babusiaux, C., , Vergely, J.~L., 
 Katz, D., Arenou, F.,  Valette, B., Hottier, C,  Capitanio, L.\  2019,
 \aap, 625, L135

\bibitem[Leike, Glatzl, \& En$\beta$lin(2020)]{Leike2020} Leike, H.,
Glatzle, M., En$\beta$lin, T.~A.\ 2020, \aap. , 639, A138

\bibitem[Linsky \& Redfield(2021)]{Linsky2021} Linsky, J.~L., Redfield, S.\  2021, \apj, 920, 75

\bibitem[Linsky et al.(2022)]{Linsky2022}  Linsky, J.~L., Redfield, S., Ryder, D., Chasan-Taber, A.\ 2022, \aj, 164, 106

\bibitem[Linsky, Rickett, and Redfield(2008)]{Linsky2008b} Linsky, J.~L.,
  Rickett, B.~J., Redfield, S.\ 2008, \apj, 675, 413

\bibitem[Ma\'iz-Apell\'aniz(2001)]{Maiz-Apellaniz2001} Ma\'iz-Apell\'aniz, J.\ 2001, \apj, 560,L83
  
\bibitem[McComas et al.(2009)]{McComas2009} McComas, D.~J., Allegrini, F., Bochsler, P.\  2009, Science, 326, 95
    
\bibitem[McComas \& Schwadron(2014)]{McComas2014} McComas, D.~J., Schwadron, N.~A.\ 2014, \apjl,  795 L17 
 
\bibitem[McKee \& Ostriker(1977)]{McKee1977} McKee, C.~F., \& Ostriker, J.~P.\ 1977, \apj, 218, 148

\bibitem[M\"uller et al.(2006)]{Muller2006} M\"uller,  H.-R., Frisch, P.~C., Florinski, V., Zank, G.~P.\  2006, \apj,  647, 1491

\bibitem[Parker(1958)]{Parker1958} Parker, E.~N.\ 1958, \apj, 128, 664

\bibitem[Parker(1961)]{Parker1961} Parker, E.~N.\ 1961, \apj, 134, 20

\bibitem[Peek et al.(2011)]{Peek2011} Peek, J.~E.~G., Heiles, C., 
Peek, K.~M.~G., Meyer, D., \& Lauroesch, J.~T.\ 2011, \apj, 735, 129

\bibitem[Pizzo(1978)]{Pizzo1978} Pizzo, V.\ 1978, \jgr, 83(A12), 5563

\bibitem[Raymond et al.(2020)]{Raymond2020} Raymond, J.~C., Slavin, J.~D.,
Blair, W.~P., et al.\ 2020, \apj, 903, 2

\bibitem[Redfield \& Linsky(2008)]{Redfield2008} Redfield. S., \& Linsky, J.~L.\ 2008, \apj, 673, 283

\bibitem[Richardson(2008)]{Richardson2008} Richardson, J.~D.\ 2008, Geophysical Research Letters, 35, L23104

\bibitem[Richardson et al.(2022)]{Richardson2022} Richardson, J.~D., Burlaga, L.~F., Elliot, H., Kurth, W.~S., Liu, Y.~D., von Steiger, R.\ 2022, \ssr, 218, 35

\bibitem[Richardson et al.(2020)]{Richardson2020} Richardson, J.~D., Belcher, J.~W., Burlaga, L.~F., Cummings, A.~C., Decker, R.~B., Opher, M., Stone, E.~C.\ 2020, J.Phys. Conf. Ser., 1620, 012016

\bibitem[Salvati(2010)]{Salvati2010} Salvati, M.\ 2010, \aap, 513, A28

\bibitem[Schwadron et al.(2011)]{Schwadron2011} Schwadron, N.~A., Allegrini, F., Bzowski, M., et al.\  2011,  \apj, 731, 56

\bibitem[Shelton(1999)]{Shelton1999} Shelton, R.~L.\ 1999, \apj, 521, 217

\bibitem[Slavin \& Frisch(2008)]{Slavin2008} Slavin, J.~D., Frisch, P.~C.\ 2008, DOI: 10.1051/0004-6361:20078101, \aap, 491, 53

\bibitem[Snowden et al.(2014)]{Snowden2014} Snowden, S.~L., Chiao, M., Collier, M.~R.\ 2014, \apj, 791, L14

\bibitem[Sobey et al.(2019)]{Sobey2019} Sobey, C.,  Bilous, A.~V., Griessmeier, J.~M.\ 2019, \mnras,  484, 3646

\bibitem[Stone et al.(2019)]{Stone2019} Stone, E.~C., Cummings, A.~C.,
  Heikkila, B.~C., Lal, N.\ 2019, Nature Astronomy, 3, 1013

\bibitem[Stone et al.(2013)]{Stone2013} Stone, E.~C., Cummings, A.~C., McDonald, F.~B., Heikkila, B.~C..,Lal, N., Webber, W.~R.\ 2013, Science, 341, 150

\bibitem[Swaczyna et al.(2016)]{Swaczyna2016} Swaczyna, P., Bzowski, M., Christian, E.~R., Funsten, H.~O., McComas, D.~J., Schwadron, N.~A.\ 2016, \apj, 823, 119

\bibitem[Swaczyna et al.(2020)]{Swaczyna2020} Swaczyna, P., McComas, D.~J., Zirnstein, E.~J., et al.\ 2020, \apj, 903, 48

\bibitem[Swaczyna et al.(2021)]{Swaczyna2021} Swaczyna, P., Rahmanifard, F., Zirnstein, E.~J., McComas, D.~J., Heerikhuisen, J.\ 2021, \apj, 911, L36 

\bibitem[Swaczyna et al.(2022b)]{Swaczyna2022b} Swaczyna, P.,  Kubiak, M.~A., Bzowski, M.\ 2022b, \apjs, 259, 42

\bibitem[Swaczyna et al.(2022a)]{Swaczyna2022a} Swaczyna, P., Schwadron, N.~A., M\"obius, E., et al.\ 2022a, \apjl, 937, L32

\bibitem[Wallner et al.(2016)]{Wallner2016} Wallner, A., Feige., Kinoshita, N., et al.\ 2016, \nat, 532, 69


\bibitem[Wolfire et al.(1995)]{Wolfire1995} Wolfire, M., McKee, C., Hollenbach, D., \& Tielens, A.\ 1995, \apj, 453, 673

\bibitem[Wolfire et al.(2003)]{Wolfire2003} Wolfire, M., McKee, C., Hollenbach, D., \& Tielens, A.~G.~G.~M.\ 2003, \apj, 587, 278


\bibitem[Zank(2015)]{Zank2015} Zank G.~P.\ 2015, \araa, 53, 449 

\bibitem[Zank \& Frisch(1999)]{Zank1999} Zank, G.~P., Frisch, P.~C.\ 1999,  \apj,  518, 965

\bibitem[Zank et al.(2013)]{Zank2013} Zank, G.~P., Heerikhuisen, J., Woods, B.~E., Pogorelov, N.~V., Zirnstein, E.~J., McComas, D.~J.\ 2013, \apj, 763, 20

\bibitem[Zirnstein et al.(2016)]{Zirnstein2016} Zirnstein, E.~J., Heerikhuisen, J., Funsten, H.~O., Livadiotis, G., McComas, D.~J.,  Pogorelov, N.~V.\ 2016, \apjl, 818, L18

\end{thebibliography}
 \end{document}